\documentclass[fleqn]{2023SCGE}
\begin{document}

\ensubject{subject}

\ArticleType{Article}
\Year{2024}
\Month{April}
\Vol{**}
\No{*}
\DOI{??}
\ArtNo{000000}
\ReceiveDate{**}
\AcceptDate{**}

\title{Forecast of cosmological constraints with superluminous supernovae from the Chinese Space Station Telescope}{Forecast of cosmological constraints with superluminous supernovae from the Chinese Space Station Telescope}

\author[1]{Xuan-Dong Jia}{}
\author[1]{Jian-Ping Hu}{}
\author[1,2]{Fa-Yin Wang}{fayinwang@nju.edu.cn}
\author[3]{Zi-Gao Dai}{}

\AuthorMark{Xuan-Dong Jia}

\AuthorCitation{X. D. Jia, J. P. Hu, F. Y. Wang and Z. G. Dai}

\address[1]{School of Astronomy and Space Science, Nanjing University, Nanjing 210093, China}
\address[2]{Key Laboratory of Modern Astronomy and Astrophysics (Nanjing University), Ministry of Education, Nanjing 210093, PR China}
\address[3]{Department of Astronomy, University of Science and Technology of China, Hefei 230026, PR China}


\abstract{
Superluminous supernovae (SLSNe) are a class of intense celestial events that can be standardized for measuring cosmological parameters, bridging the gap between type Ia supernovae and the cosmic microwave background. In this work, we  discuss the cosmological applications of SLSNe from the Chinese Space Station Telescope (CSST). Our estimation suggests that SLSNe rate is biased tracing the cosmic star formation rate, exhibiting a factor of $(1+z)^{1.2}$. We futher predict that CSST is poised to observe $\sim 360$ SLSNe in the 10 square degrees ultra-deep field survey within a span of 2.5 years. A stringent constraint on cosmological parameters can be derived from their peak-color relationship. CSST is anticipated to uncover a substantial number of SLSNe, contributing to a deeper understanding of their central engines and shedding light on the nature of dark energy at high redshifts.
}



\maketitle


\begin{multicols}{2}
\section{Introduction}\label{Introduction}

As standard candles, type Ia supernovae (SNe Ia) have provided compelling evidence for cosmic acceleration \cite{1998AJ....116.1009R,1999ApJ...517..565P,2016ApJ...826...56R,2018ApJ...853..126R}. An exotic form of energy namely, the dark energy, which occupies about $70$ per cent of the energy density at present has been purposed to account for such accelerated expansion of the universe. Although the standard cosmological-constant $\Lambda$ cold dark matter (CDM) model is remarkably successful in explaining most cosmological observations \cite{2020A&A...641A...6P,2021MNRAS.504.2535I}, there are still critical issues that remain unreconciled. \cite{1989RvMP...61....1W,2022NewAR..9501659P}. One of the most serious challenges is the Hubble tension \cite{2019NatAs...3..891V,2023Univ....9...94H,Verde2023}. Based on the local distance ladder,  the SH0ES team derives the Hubble constant of \Authorfootnote $H_{0,z\sim0} = 73.04\pm 1.04$ km s$^{-1}$ Mpc$^{-1}$ \cite{2022ApJ...934L...7R}, while the $Planck$ cosmic microwave background (CMB) anisotropies measurements extrapolates $H_{0,z\sim1100} = 67.4\pm 0.5$ km s$^{-1}$ Mpc$^{-1}$ with the $\Lambda$CDM model assumption \cite{2020A&A...641A...6P}. The discrepancy between these two approaches exceeds $5 \sigma$ \cite{2022ApJ...934L...7R}. On the way of solving the Hubble tension, many theoretical models have been proposed \cite{2021CQGra..38o3001D,2021A&ARv..29....9S}.  Recently, growing evidence for the redshift dependence of the Hubble constant has been reported \cite{Wong2020,2020PhRvD.102b3520K,Krishnan2021,Dainotti2021,2022MNRAS.517..576H,2023A&A...674A..45J}. Howerer, SNe Ia turn off at higher redshifts due to the delay time required to form a white dwarf. Even for the most distant SNe Ia discovered at a redshift of z$\sim2.26$, there is still a huge gap between SNe Ia and CMB ($z\sim$1100, \cite{2022ApJ...938..113S,2020A&A...641A...6P}.) Opportunities to explore the cosmic blank history may be provided by energetic beacons shining at the distant universe, including gamma-ray bursts \cite{Wang2015,2021MNRAS.507..730H,2021JCAP...09..042K,Wang2022,2022MNRAS.516.1386C}, quasars \cite{Risaliti2019,2022MNRAS.513.5686C,2022MNRAS.516.1721C} and SLSNe.

Superluminous supernovae (SLSNe) belong to a subtype of extremely energetic astronomical transients. Since the discovery of SN\,1999as, substantial efforts to search and followup observations have revealed such an exotic class of cosmic explosions, which are characterized by their extraordinary peak magnitudes, typically with $M \leq -21$ mag and a remarkably slow evolution compared to other known types of SNe, with a typical rising time of $\sim$50--100 days in the rest frame \cite{1999IAUC.7128....1K}. Their peak luminosities  may surpass that of normal  SNe Ia by a factor of 100, reaching up to $10^{43-44}$ erg s$^{-1}$, while the total radiated energy amounts to $\sim 10^{51}$ erg. ({\textbf see {\cite{2019ARA&A..57..305G} for a review}}).
Based on their photometric and spectroscopic properties, SLSNe can be divided into two subclasses, namely Types I and II , which have been characterized by the absence and presence of hydrogen lines, respectively. Hydrogen features are hereby indicative of the collision between the ejecta and its circumstellar matter (CSM, \cite{2007ApJ...666.1116S,2010ApJ...718L.127D,2011ApJ...729..143C,2014MNRAS.441..289B}). Energy sources that power the light curve of SLSNe remain unclear. Promising mechanisms include, (i) the rapid spin-down of a magnetar, which accounts well for the near-peak luminosity evolution \cite{2010ApJ...717..245K,2010ApJ...719L.204W,2016ApJ...817..132D,2017ApJ...840...12Y,2020MNRAS.497..318L}, (ii) the radioactive decay of up to several tens of solar masses of freshly nucleo-synthesized nickel \cite{2002ApJ...567..532H,2003ApJ...591..288H}, and (iii) the interaction of the SN ejecta and its surrounding CSM, implying episodes of mass-loss of the progenitor before its terminate explosion \cite{Ginzburg2012,2022A&A...666A..30P,2023NatAs...7..779L}.

The remarkably high luminosity of SLSNe allows them to be observed at higher redshifts, with the currently highest recorded redshift reaching up to $z \sim 4$ \cite{2012Natur.491..228C,2013ApJ...779...98H}. This distinctive feature has triggered great interest in using them as cosmological distance indicators. For instance, Quimby et al. \cite{2013MNRAS.431..912Q} was the first one who showed SLSNe might be a standard candle.
A sample of 16 SLSNe I, spanning the redshift range $0.1 < z < 1.4$, was employed to develop a standardization method \cite{2014ApJ...796...87I}. Despite the limited sample size, the results tentatively suggest a correlation between the peak magnitude and the decline rate.
Until recently, by incorporating a peak magnitude - color evolution found for SLSNe, such a peak magnitude - decline rate relation has been utilized to construct a Hubble diagram that extends to redshifts up to $z\sim$2 \cite{2021MNRAS.504.2535I}, which is beyond the capabilities of SNe Ia. These findings underscore the potential of SLSNe as standard candles for constraining cosmological parameters. 
Nevertheless, the current  sample of SLSNe for cosmological research remain scare, which could be substantially enriched by the Chinese Space Station Telescope (CSST) \cite{2011SSPMA..41.1441Z}.

CSST is a sizable space-based astronomical telescope initiated under the China Manned Space Program. 
With a planned launch and the commencement of scientific operations set for 2024, this 2-meter aperture telescope will share the orbit of the China Manned Space Station, completing an orbit around the Earth approximately every 90 minutes. More detailed descriptions of the imaging and slitless spectroscopic capabilities of CSST can be found in earlier literature \cite{2018MNRAS.480.2178C,2019ApJ...883..203G}. The optical and orbital design of CSST yields a total survey area of $\gtrsim$30,000 deg$^{2}$ \cite{2011SSPMA..41.1441Z}. Although the orientation of the space station telescope will be affected by its anchoring space station, thus limiting its capability to carry out continuous and prolonged integrations on the same sky area, it can be compensated, to some extent, by conducting multiple sets of short exposures on the same area and stacking the images later on.

Multiple sky survey modes will be used to achieve various science goals, including a large sky area multicolor imaging survey, a large sky area slitless spectroscopic survey, an ultra-deep multicolor imaging survey, and provisions for flexible scheduling to observe unpredictable phenomena that are of high scientific merit. The exposure time for the primary survey mission is nominally 150 seconds, while the maximum exposure time can reach up to 300 seconds. The $5\sigma$ limiting AB magnitudes of the telescope in imaging mode give NUV $\approx 25.4$, u $\approx 25.4$, g $\approx 26.3$, r $\approx 26.0$, i $\approx 25.9$, z $\approx 25.2$, and y $\approx 24.4$. Further details are presented in \cite{2023SCPMA..6629511L}. To search for and describe the physical properties of the dark energy is one of the major scientific goals of the CSST \cite{2011SSPMA..41.1441Z}. 

The high spatial resolution ($\sim 0.^{\prime \prime}15$) combined with the large field of view (1.1 deg$^2$, \cite{2019ApJ...883..203G}) enables CSST a promising facility in discovering and monitoring SLSNe. These slowly-evolving transients, i.e. iPTF13ajg \cite{2013Natur.502..346N}, with a characteristic rest-frame rise time of 50-100 days and further prolonged by a factor of $(1+z)$, can still be comprehensively sampled even with a relatively low cadence of 10 days under the 2.5-year long, ultra-deep survey mode of CSST. 
The long time-baseline, multi-color monitoring of all SLSNe within the CSST survey footprint will also offer insights into the physical properties of SLSNe and enable diagnostics of various models of its energy sources. To detect SLSNe, an ultra-deep survey program covering $10$ square degrees is proposed as a supplement to the major survey.

This paper investigates the capability of CSST to discover Type I SLSNe in the proposed 10 deg$^2$ ultra-deep survey sky area. Given the detectability of SLSNe at high redshifts, we employ a simulated CSST sample to constrain cosmological parameters. The process of calculating the rate of SLSNe is detailed in Section \ref{The Rate Of SLSNe}. Section \ref{Standardisable Candle} outlines the method employed for standardizing SLSNe, while the peak-color relation is refitted. It is subsequently utilized to generate simulated SLSNe observed by CSST. The conclusions are summarized in Section \ref{Conclusions}.

\section{The Rate of SLSNe}\label{The Rate Of SLSNe}
\subsection{The ZTF Sample of SLSNe-I}
Being the successor of the Intermediate Palomar Transient Factory (iPTF, \cite{2009PASP..121.1395L}), the Zwicky Transient Facility employs a 600 egapixel camera mounted on the Palomar Samuel Oschin $48$ inch Schmidt telescope, enabling it to achieve an large field of view of $47$ deg$^2$ \cite{2019PASP..131a8002B,2019PASP..131g8001G,2019PASP..131a8003M,2020PASP..132c8001D}. ZTF scans the sky with unprecedented efficiency at a depth of 20.8~magnitude. It can cover the full northern sky in 3 days, which is particularly well-suited for the detection of transient events. For example, the Northern Sky Public Survey observes the entire accessible northern sky, covering $\approx$23.675 deg$^{2}$ with a $\sim3$ day cadence. Within the footprint of ZTF Phase I, a sample of $85$ SLSNe-I have been discovered and spectroscopically confirmed between $2018$ March $17$ and $2020$ October $31$ This sample of ZTF SLSNe-I spans a redshift range of $z = 0.06 - 0.67$ \cite{2023ApJ...943...41C}. The peak luminosity covers -22.8 mag $\le M_{peak} \le$ -19.8 mag, with a median value of $-21.48^{+1.13}_{-0.61}$. We here use the luminosity function of the SLSNe-I based on the relatively large sample of spectroscopically confirmed SLSNe-I collected by ZTF.

\subsection{Calculating the rate of SLSNe}
The determination of volumetric rates for SLSNe challenging primarily due to limited sample sizes \cite{2013MNRAS.431..912Q,2015MNRAS.448.1206M,2017MNRAS.464.3568P,2021MNRAS.500.5142F}. The details of the methodology are outlined below.

First, we determine the maximum redshift out to which the telescope could observe SLSNe. It is determined by the maximum brightness exhibited by the SLSNe sample and the observational limitations imposed by the telescope. 
The volumetric rate of SLSNe is defined as a sum over $N$ SLSNe events found in a given comoving volume $V$ during $T$ years of observations, namely:
\begin{equation}
    r_{\mathrm{SLSNe}}=\frac{1}{V} \sum_{i}^{N} \frac{\left(1+z_{i}\right)}{\epsilon_{i} T_{i}}.
\end{equation}
The factor $(1+z_i)$ corrects for the time dilation. The detection efficiency $\epsilon$ denotes the ratio between the SLSNe identified and covered by the footprint of the survey. The comoving volume element $V$ is
\begin{equation}
	V=\frac{\Theta}{41253} \frac{4 \pi}{3}\left[\frac{c}{H_{0}} \int_{0}^{z} \frac{\mathrm{d} z^{\prime}}{\sqrt{\left.\Omega_{m}\left(1+z^{\prime}\right)^{3}+\Omega_{\Lambda}\right)}}\right]^{3} \mathrm{Gpc^3},
\end{equation}
where $\Theta$ represents the survey area in units of deg$^{2}$ and c gives the speed of light in units of km\,s$^{-1}$. Cosmological parameters are assumed to fiducial values, $\Omega_m = 0.3$, $\Omega_{\Lambda} = 0.7$ and $H_0 = 73$ km s$^{-1}$ Mpc$^{-1}$. 

We can predict the number of observations based on the event rate $r_{\mathrm{SLSNe}}$ and the comoving volume $V$. The method has been used since the prediction of $Euclid$ \cite{2021MNRAS.500.5142F,2022A&A...666A.157M}. In the next section, we calculate the volumetric rate with redshift and predict the corresponding number of SLSNe observed by CSST.

\subsection{Redshift dependence of the SLSN rate}
As SLSNe are associated with the death of very massive stars, their rate would be considered to be roughly proportional to the specific star-formation rate (SFR normalized by the stellar mass of the host galaxy, see, e.g., \cite{2007A&A...468...33E,2021ApJS..255...29S}) of the host galaxy. Substantial effort has gone into the estimation of the volumetric rate for SLSNe. However, the uncertainty remains high due to the limited sample size of SLSNe.

Based on a single event of SN\,2005ap in the Robotic Optical Transient Search Experiment-IIIb telescope, a volumetric rate of SLSNe at redshift z=0.17 has been estimated to be about $32^{+77}_{-26}$ Gpc$^{-3}$yr$^{-1}$ \cite{2013MNRAS.431..912Q}. Discovery of the two candidate SLSNe discovered by the Pan-STARRS Medium Deep Survey, namely PS1-10pn and PS1-10ahf, suggests that the rate of SLSNe would between $3^{+3}_{-2} \times 10^{-5}$ and $8^{+2}_{-1} \times 10^{-5}$ that of the core-collapse SNe rate over the redshift range $0.3 \sim 1.4$ \cite{2015MNRAS.448.1206M}. In a more recent research, Prajs et al. \cite{2017MNRAS.464.3568P} used three SLSNe from the Canada-France Hawaii Telescope Supernova Legacy Survey to find that $r_{\mathrm{SLSNe}} = 91^{+76}_{-36}$ Gpc$^{-3}$yr$^{-1}$ at a weighted mean redshift of $z = 1.13$. From the PTF data sample, Frohmaier et al. \cite{2021MNRAS.500.5142F} used eight events to derive a volumetric rate for SLSNe as $r_{\mathrm{SLSNe}} = 35^{+25}_{-13}$ Gpc$^{-3}$ yr$^{-1}$ at $z \leq 0.2$. The estimation of the highest redshift presents an optimistic rate of $\sim 400$ Gpc$^{-3}$ yr$^{-1}$ at a weighted redshift of $z = 3$ based on two SLSNe \cite{2012Natur.491..228C}. 

\begin{figure}[H]
\centering
\includegraphics[width=0.47\textwidth,angle=0]{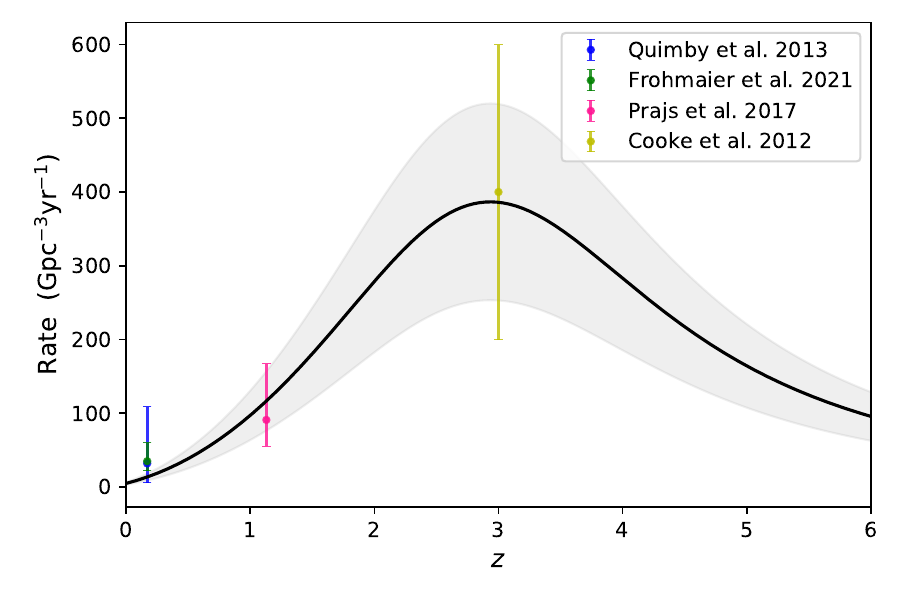}
\caption{The volumetric rate of SLSNe as a function of redshift. The black line and light gray region represent the volumetric rate of SLSNe and its associated 1$\sigma$ uncertainty, respectively.\label{F_Rate}}
\end{figure}

Referring to the predictions of $Euclid$ and the calibration established in prior literature \cite{2012Natur.491..228C,2013MNRAS.431..912Q,2015MNRAS.448.1206M,2017MNRAS.464.3568P,2018A&A...609A..83I,2021MNRAS.500.5142F,2022A&A...666A.157M}, here we derive a form of the SLSNe rate. To investigate the redshift dependence of $r_{\rm SLSNe}$, following the form that describes the space density of the SFR as a function of its initial mass function (IMF, \cite{2001MNRAS.326..255C}) and the parameterization given in the prior literature \cite{2006ApJ...651..142H}, we posit a linear dependence on the SFR with a factor of $(1+z)^{\beta}$. The rate of SLSNe is expressed as follows
\begin{equation}
	r_{\mathrm{SLSNe}}=R_{0} (1+z)^{\beta} \times \mathrm{SFR}(z) ~\mathrm{Gpc}^{-3}\mathrm{yr}^{-1}  ,
\end{equation}
where $R_0$ is the correlation coefficient. The efficiency of SLSNe production per unit stellar mass is expressed as an empirical form $\epsilon (z) = R_{0} (1+z)^{\beta}$, which follows that a format akin to the efficiency of gamma-ray bursts \cite{Wang2013,2013ApJ...773L..22T}.
Adopting the SLSNe rates derived by previous studies \cite{2012Natur.491..228C,2013MNRAS.431..912Q,2017MNRAS.464.3568P,2021MNRAS.500.5142F}, we infer the coefficient $R_0 = 411.64 \pm 142$ and $\beta=1.2$. Since there are only four known event rates currently, which can not sufficiently constrain the evolution, we fixed the exponent $\beta$ ranging from 0 to 1.5, and then use Markov Chain Monte Carlo (MCMC) to fit the coefficient $R_0$. The volumetric rate of SLSNe as a function of redshift is shown in Fig. \ref{F_Rate}. The uncertainty range of the volumetric rates is calculated from the uncertainties of $R_0$. The volumetric rates of SLSNe in previous literature are consistent with our best-fit solid line within $1 \sigma$ uncertainties.

\begin{figure}[H]
\centering
\includegraphics[width=0.47\textwidth,angle=0]{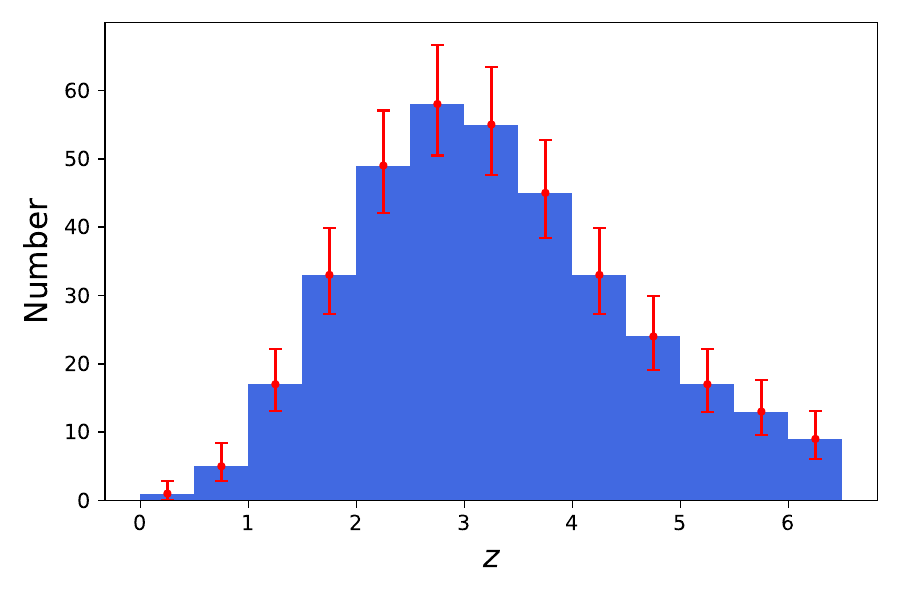}
\caption{Number density distribution of SLSNe to be observed by the 2.5-year long, 10\,deg$^{2}$ CSST ultra-deep sky survey. The size of the error bars represents the 1$\sigma$ Poisson uncertainties estimated within each redshift bin.}
\label{F_Distribution}
\end{figure}

Following the prescriptions in \cite{2018A&A...609A..83I}, we use $r_{\mathrm{SLSNe}}$ to calculate the expected number of SLSNe to be observed by CSST. The redshift bin is adopted as $\Delta z = 0.5$. The number of SLSNe within each $\Delta z$ is determined by the mean value of SLSNe rate at that bin and the corresponding volume. The distribution of the SLSNe sample that will be observed by CSST is shown in Fig. \ref{F_Distribution}. The error bars in the histograms represent the 1$\sigma$ Poission uncertainties determined from the count in each redshift bin \cite{1986ApJ...303..336G}. More details about the calculation will be given in Section \ref{Data Simulation}.

\section{Standardisable Candle}\label{Standardisable Candle}
\subsection{SLSNe sample}
The sample of SLSNe utilized in this paper has been compiled from previous literature \cite{2021MNRAS.504.2535I}. As shown in Table \ref{SLSNe data}, it consists of the observations from the Panoramic Survey Telescope and Rapid Response System 1 (PanSTARRS-1) \cite{2018ApJ...852...81L}, the Palomar Transient Factory (PTF) and intermediate PTF (iPTF) \cite{2018ApJ...860..100D} and Dark Energy Survey (DES) \cite{2019MNRAS.487.2215A}. The candidates which pass the data filtering steps have (i) a light curve covering -15 to +30 d in the rest frame and without multiple peaks, (ii) a spectrum between -15 and +30 d in the rest frame and (iii) a SLSNe I likely hypersurface defined by four photometric variables (Four Obervables Parameter Space, \cite{2021MNRAS.504.2535I}). The SLSNe-I category can be further divided into two distinct subclasses according to how fast their near-peak light curves evolve, namely the fast and slow \cite{2018ApJ...854..175I,2018ApJ...855....2Q,2019ARA&A..57..305G,2019NatAs...3..697I}. The former have a light-curve evolution similar to SN 2010gx \cite{2010ApJ...724L..16P}. The fast subclasses constitute the majority of the SLSNe population and display a tight `peak-decline' relation. The other subclass with objects similar to SN 2007bi does not do well on this relation \cite{2009Natur.462..624G,2010A&A...512A..70Y}. Such slowly-evolving events show signatures of interactions between the ejecta and the ambient circumstellar medium \cite{2016ApJ...826...39N,2017MNRAS.468.4642I,2017ApJ...848....6Y,2019NatAs...3..697I}. The SLSNe sample comprises 20 SLSNe, with 15 falling into the Fast subclass, and the remaining 5 categorized as `unclassified' due to the absence of classification spectra. We remark that for the sake of homogeneity of the sample, we only adopt the cases belonging to the `fast' subclass, whose light-curve and spectrophotometric behavior indicates to be homogeneous.

The utilization of SLSNe as standard candles has been extensively explored \cite{2014ApJ...796...87I,2021MNRAS.504.2535I}. It has been demonstrated that SLSNe are capable of constraining cosmological parameters. Parameters that have been tested to be potentially useful to establish correlations include the peak luminosity in the $400$ nm filter ($M(400)_0$), and the corresponding magnitude decline over the $30$ days from the peak brightness ($\Delta M(400)_{30}$), and its difference with that measured in the 520 nm filter ($M(400)_{30}-M(520)_{30}$). As shown in Table \ref{SLSNe data}, $M(400)_{30}$, $\Delta M(400)_{30}$, and $M(400)_{30}-M(520)_{30}$ are available for 20, 20, and 14 SLSNe, respectively.

\subsection{SLSNe Standardization}\label{SLSN I Standardization}
There are two known pairs of parameters to standardize SLSNe, namely the `peak-decline' ($M(400)_{0}$ - $\Delta M(400)_{30}$) and the `peak-colour' ($M(400)_{0}$ - $M(400)_{30}-M(520)_{30}$) correlations, both can be expressed as a linear form, i.e.,

\begin{equation}
y = a x + b.
\end{equation}
The independent variable $x$ refers to $\Delta M(400)_{30}$ and $M(400)_{30}-M(520)_{30}$ for the `peak-decline' and the `peak-colour' correlations, respectively. The dependent variable $y$ means the peak luminosity $M(400)_0$.

\begin{figure}[H]
\centering
\includegraphics[width=0.45\textwidth,angle=0]{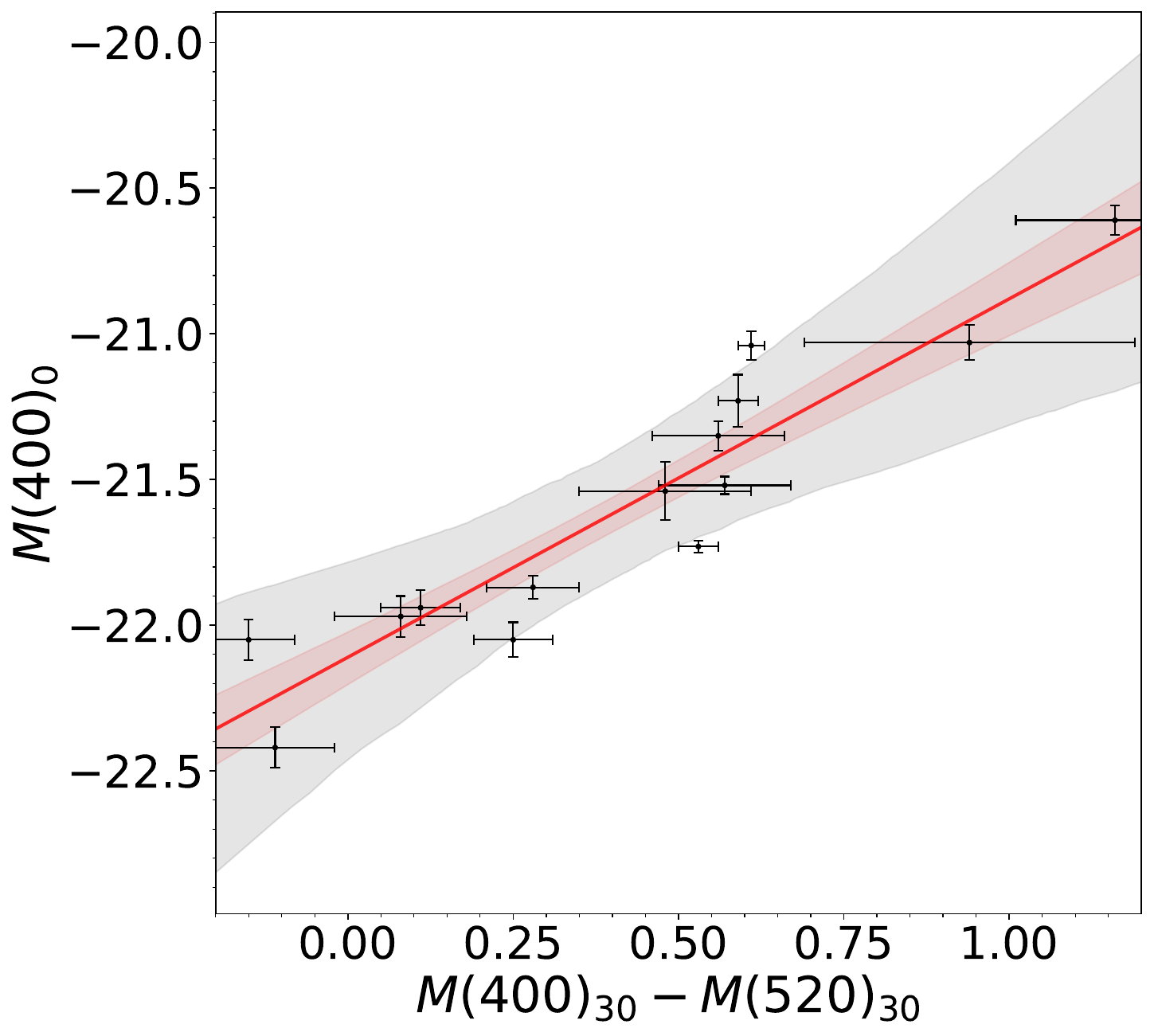}
\caption{The $M(400)_{0}$ - $M(400)_{30}-M(520)_{30}$ relation derived from the observations of 14 SLSNe. The solid red line represents the best fit. The light red region and the light black region represent the 1$-$ and 2$-\sigma$ uncertainties, respectively.}
\label{F_M0M45}
\end{figure}

Our fitting procedure follows a MCMC approach by using the emcee\footnote{\url{https://emcee.readthedocs.io/en/stable/}} package \cite{2013PASP..125..306F}. The likelihood function is 
\begin{equation}
\begin{aligned}
\mathcal{L}\left( a, b, \sigma_{\mathrm{ext}}\right) \propto \prod_{i} &   \frac{1}{\sqrt{\sigma_{\mathrm{ext}}^{2}+\sigma_{y_{i}}^{2}+a^{2} \sigma_{x_{i}}^{2}}} \\
& \times \exp \left[-\frac{\left(y_{i}-a x_{i}-b \right)^{2}}{2\left(\sigma_{\mathrm{ext}}^{2}+\sigma_{y_{i}}^{2}+a^{2} \sigma_{x_{i}}^{2}\right)}\right] ,
\end{aligned}
\end{equation}
where $x_i$ and $y_i$ are the data for the $i$th SLSNe. The best-fit values with $1\sigma$ uncertainties are $a = 0.75 \pm 0.26$, $b = -22.32 \pm 0.23$ and $\sigma_{\mathrm{ext}} = 0.44 \pm 0.09$ for the $M(400)_{0}$ - $\Delta M(400)_{30}$ relation, and $a = 1.23 \pm 0.19$, $b = -22.11 \pm 0.10$ and $\sigma_{\mathrm{ext}} = 0.20 \pm 0.06$ for the $M(400)_{0}$ - $M(400)_{30}-M(520)_{30}$ relation. The latter, which is illustrated in Fig. \ref{F_M0M45}, yields a tighter correlation compared to the former as indicated by the overall smaller uncertainties in the fitted parameters.

Our MCMC fitting to both relationships discussed above results are consistent with those reported in previous literature within their 1$-\sigma$ uncertainties \cite{2021MNRAS.504.2535I}. It is worth mentioning that the values of the absolute magnitudes are calculated from the rest-frame apparent magnitudes with a same cosmology assumption ($H_0 = 72$ km s$^{-1}$ Mpc$^{-1}$, $\Omega_m = 0.27$, $\Omega_{\Lambda} = 0.73$). Therefore, restricting the cosmological model directly with the well-fitted relations will result in circular argumentation. The main purpose of the above analysis is to justify

\subsection{Simulating the CSST SLSNe sample}\label{Data Simulation}
We hereby simulate a sample of SLSNe that will be observed by the 2.5-year, 10\,deg$^{2}$ CSST ultra-deep survey. The exposure time for the survey mission of the CSST is nominally $150$ s. Considering the effective wavelength and limiting magnitudes for point sources detected at $5 \sigma$, our simulation extends to a redshift of $z\approx$6.5, compared to $z\approx4$ for the most distant known SLSNe \cite{2012Natur.491..228C}. Based on the volumetric rate of SLSNe discussed above, our estimation suggests that a total number of 360 events will be identified. The methodology of the simulation is briefly outlined below.

Firstly, we adopt the rather tighter $M(400)_{0}$ - $M(400)_{30}-M(520)_{30}$ relation, and generate a series of simulated $\Delta M(400)_{30}$ based on the observed sample. The range of the generated simulation data falls within the observational data range. The the mock series of $\Delta M(400)_{30}$ has been assumed to be uniformly distributed within the range of observed sample, which is consistent with that seen from the sample of 14 SLSNe.

Secondly, we build upon the peak-colour relation. The scattering of $\Delta M(400)_{0}$ caused by the deviation of the simulated ($M_{\mathrm{sim}}(400)_{0}$) and the fiducial $M_{(400)_{0}}$ yields
\begin{equation}
    M_{\mathrm{sim}}(400)_0 = \alpha \Delta M_{\mathrm{sim}}(400)_{30} + \beta + \Delta M(400)_0 .
\end{equation}
Where $\alpha$ and $\beta$ are determined from Section \ref{SLSN I Standardization}. The fiducial $M(400)_0$ is calculated as $M(400)_0 = \alpha \Delta M_{\mathrm{sim}}(400)_{30} + \beta$. The deviation $\Delta M(400)_0$ is assumed to obey a Gaussian distribution $\mathcal N (0 , \sigma_{\mathrm{sim}})$. 

We then filter the simulated data with the CSST observational constraints.
At a given redshift, the apparent magnitude of the SLSNe should fall within the observation limit of CSST. We calculate the former based on the cosmological parameters $H_0 = 73$ km s$^{-1}$ Mpc$^{-1}$, $\Omega_m = 0.3$, and $\Omega_{\Lambda} = 0.7$. The distribution of the results with respect to redshifts is illustrated in Fig. \ref{F_Distribution}.

\begin{figure}[H]
	\centering
	\includegraphics[width=0.5\textwidth,angle=0]{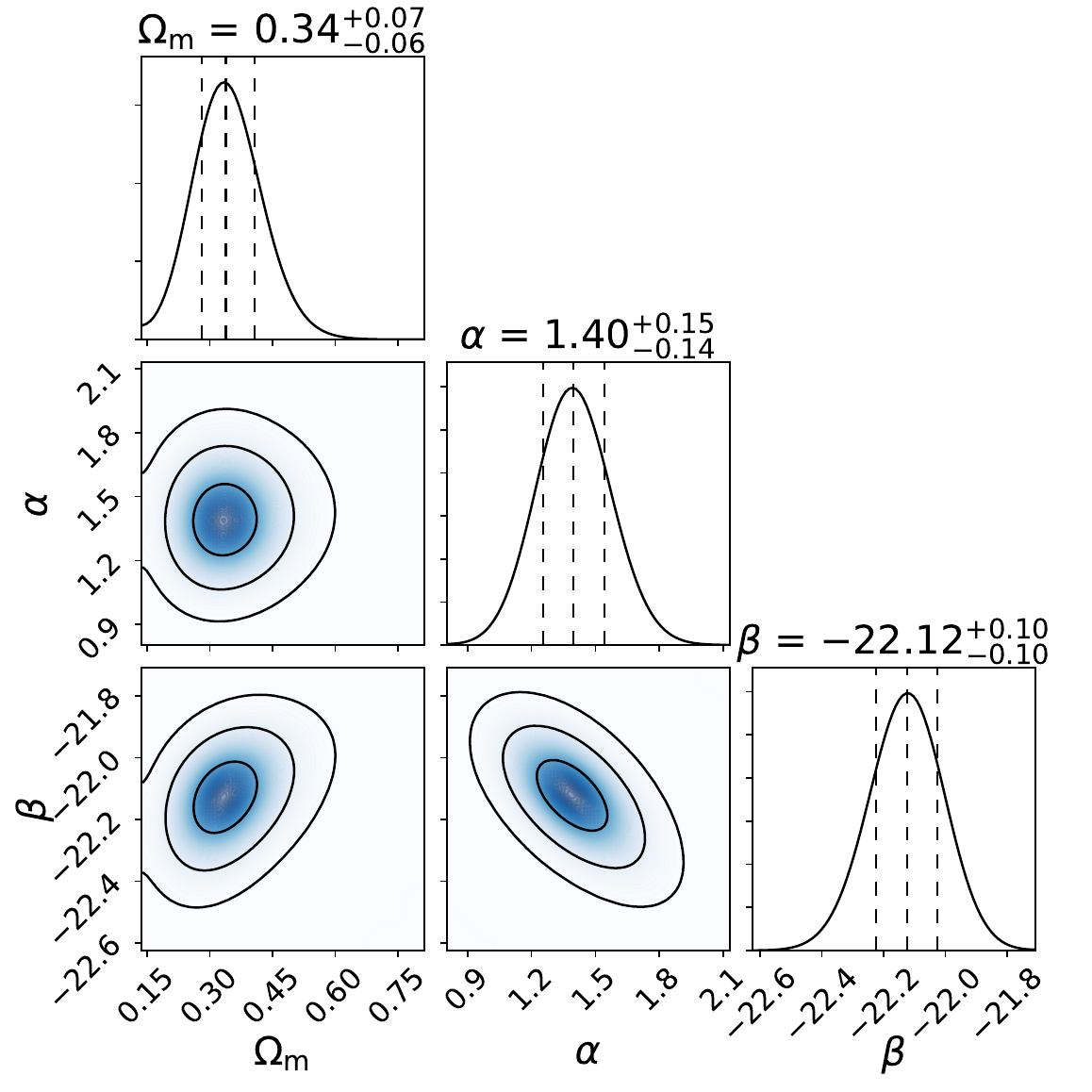}
	\caption{Joint confidence level contours of the $\Lambda$CDM model parameters inferred from the MCMC-based fitting of the SLSNe sample. The inner, middle, and outer contours centered at any intersection of parameter pairs indicate the $68\%, 95\%$, and $99.7\%$ confidence levels, respectively.
	}
	\label{F_LCDM}
\end{figure}

\subsection{Cosmology models with SLSNe}
Based on the compilation of previous observed SLSNe which provides up-to-date statistics, our simulated sample enables the most up-to-date estimation of the interpretive power of SLSNe cosmology. Here, we use the peak-colour relation to standardized them in the cosmological analysis. Our calculation assumes that SLSNe with an identical colour decline rate possess the same absolute magnitude at all redshifts. 
The standardized distance modulus ($\mu_{\mathrm{obs}}$) is then given by
\begin{equation}
    \mu_{\mathrm{obs}} = m(400) - \alpha \left[M(400)_{30}-M(520)_{30}\right] - \beta,
\end{equation}
which will be used to compare with the model distance modulus ($\mu_{\mathrm{model}}$).  According to the result from {\it Planck} CMB measurements \cite{2020A&A...641A...6P}, the space is extremely flat, which gives $\Omega_{\mathrm{k}}=0$. The cosmological models considered in this analysis are limited to the Flat $\Lambda$CDM and the Flat $w$CDM model. For the flat $\Lambda$CDM model and $w$CDM models, the Hubble parameter gives:
\begin{equation}
    H(z)_{\Lambda \rm{CDM}}=H_{0} \sqrt{\Omega_{m}(1+z)^{3}+\Omega_{\Lambda}}.
\end{equation}
and
\begin{equation}
     H(z)_{w \rm{CDM}}=H_{0} \sqrt{\Omega_{m}(1+z)^{3}+\Omega_{\mathrm{DE}}(1+z)^{3(1+w)}},
\end{equation}
respectively. Where $\Omega_{\mathrm{DE}}$ is the dark energy density parameter and $w$ is the dark energy equation of state parameter.

\begin{figure}[H]
	\centering
	\includegraphics[width=0.5\textwidth,angle=0]{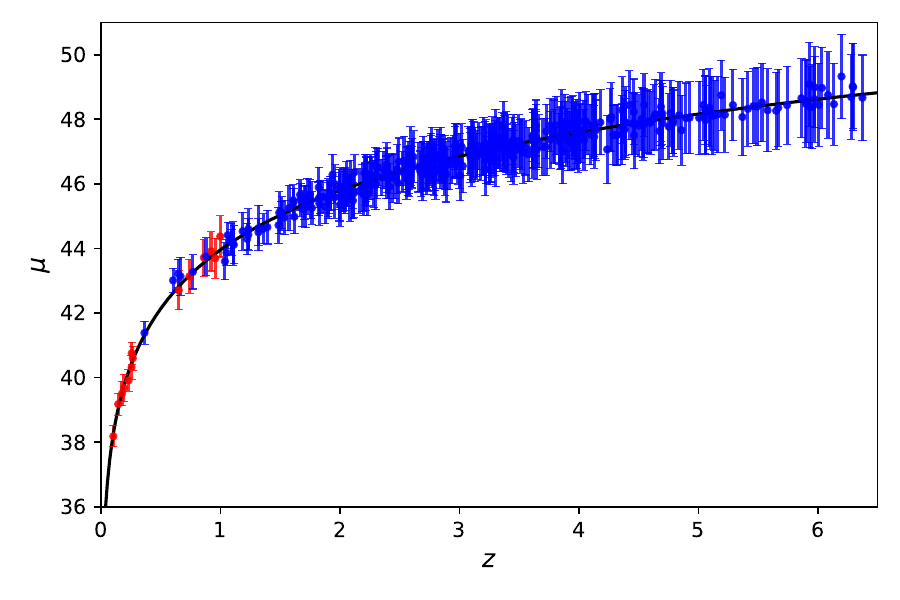}
	\caption{Hubble diagram of the observed and predicted SLSNe. The red and blue dots represent the observed cases and the model-predicted events, respectively.}
	\label{F_Modulus}
\end{figure}

The process of fitting the parameters is then minimizing the $\chi^2$ according to 
\begin{equation}
\chi_{\mathrm{SLSNe}}^{2}=\sum_{i=1}^{N} \frac{\left[\mu_{\mathrm{obs}}\left(z_{i}\right)-\mu_{\mathrm{model}}\left(z_{i}\right)\right]^{2}}{\sigma^{2}_\mu (z_{i})}.
\end{equation}
Note that the analysis is dominated by statistical uncertainties due to the small sample size. Systematic errors are not considered here. The uncertainty in the distance modulus has been estimated following the description given in \cite{2021MNRAS.504.2535I}. The Hubble constant ($H_0$) can not be constrained by SLSNe alone because it appears in both $\mu_{\mathrm{obs}}$ and $\mu_{\mathrm{model}}$. Therefore, additional probes such as SNe Ia need to be incorporated to determine $H_0$. Under the cosmological models mentioned above, the free parameters in this fit are $\Omega_m$, $w$, $H_0$, $\alpha$ and $\beta$ in a flat universe. 

For the MCMC analysis, the priors used for parameters are: $\Omega_{m} \in$ [0,1], $w \in$ [-3,0.33], $H_0 \in$ [50,80], $\alpha \in$ [-5,5] and $\beta \in$ [-30,-20]. The results indicate that the SLSNe sample effectively constrains cosmological parameters. To obtain more precise constraints on the Hubble constant $H_0$, we further combine them with the Pantheon+ SNe Ia sample. The latter consists of $1701$ light curves of 1550 SNe\,Ia, where were all adopted in our analysis \cite{2022ApJ...938..113S}.

\begin{figure}[H]
\centering
\includegraphics[width=0.5\textwidth,angle=0]{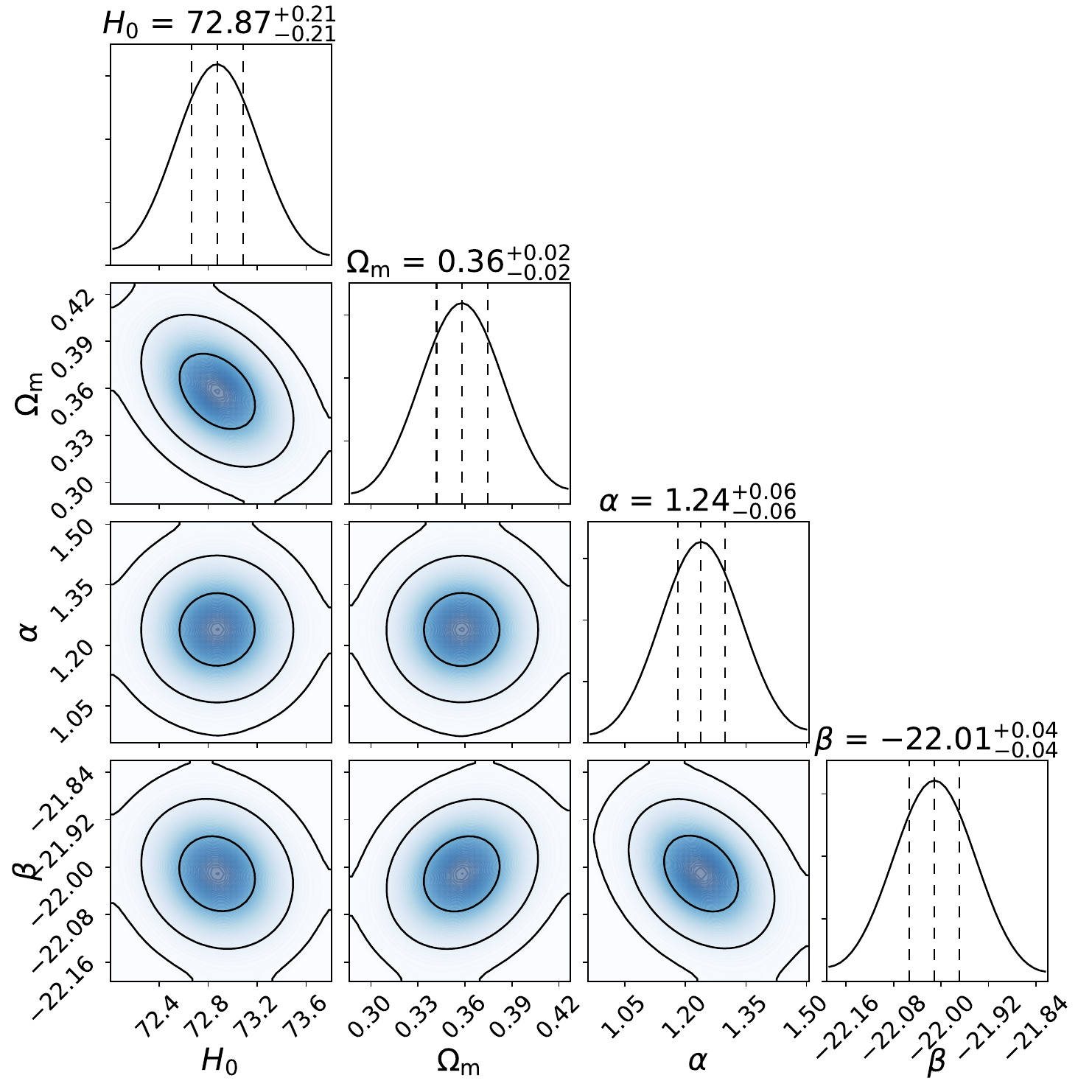}
\caption{Joint confidence level contours of the $\Lambda$CDM model parameters inferred from the MCMC-based fitting of the SLSNe and SNe sample. The inner, middle, and outer contours centered at any intersection of parameter pairs indicate the $68\%, 95\%$, and $99.7\%$ confidence levels, respectively.
}
\label{F_LCDM_SNe}
\end{figure}

As presented in Fig. \ref{F_LCDM}, for the flat $\Lambda$CDM model, the best-fit results are $\Omega_m = 0.34^{+0.07}_{-0.06}$, $\alpha = 1.40^{+0.15}_{-0.14}$ and $\beta = -22.12 \pm 0.10$. The results are consistent with those from \cite{2021MNRAS.504.2535I} within $1 \sigma$ uncertainties. The corner panel shows the best-fit values and $1 \sigma$ uncertainties. The associated Hubble diagram is shown in Fig. \ref{F_Modulus}. The red points mark the observed SLSNe, and the blue points show the simulated SLSNe. The black line is the fiducial cosmology model with $\Omega_m = 0.34$ and $H_0 = 73$ km s$^{-1}$ Mpc$^{-1}$. A more refined constraint is achieved by combining the SLSNe sample with the Pantheon+ sample. The results are $H_0 = 73.87 \pm 0.21$ km s$^{-1}$ Mpc$^{-1}$, $\Omega_m = 0.36 \pm 0.02$, $\alpha = 1.24 \pm 0.06$ and $\beta = -22.01 \pm 0.04$. The confidence regions of results are shown in Fig. \ref{F_LCDM_SNe}. They are consistent with the results from \cite{2022ApJ...934L...7R} within their 1$-\sigma$ uncertainties.

\begin{figure}[H]
\centering
\includegraphics[width=0.5\textwidth,angle=0]{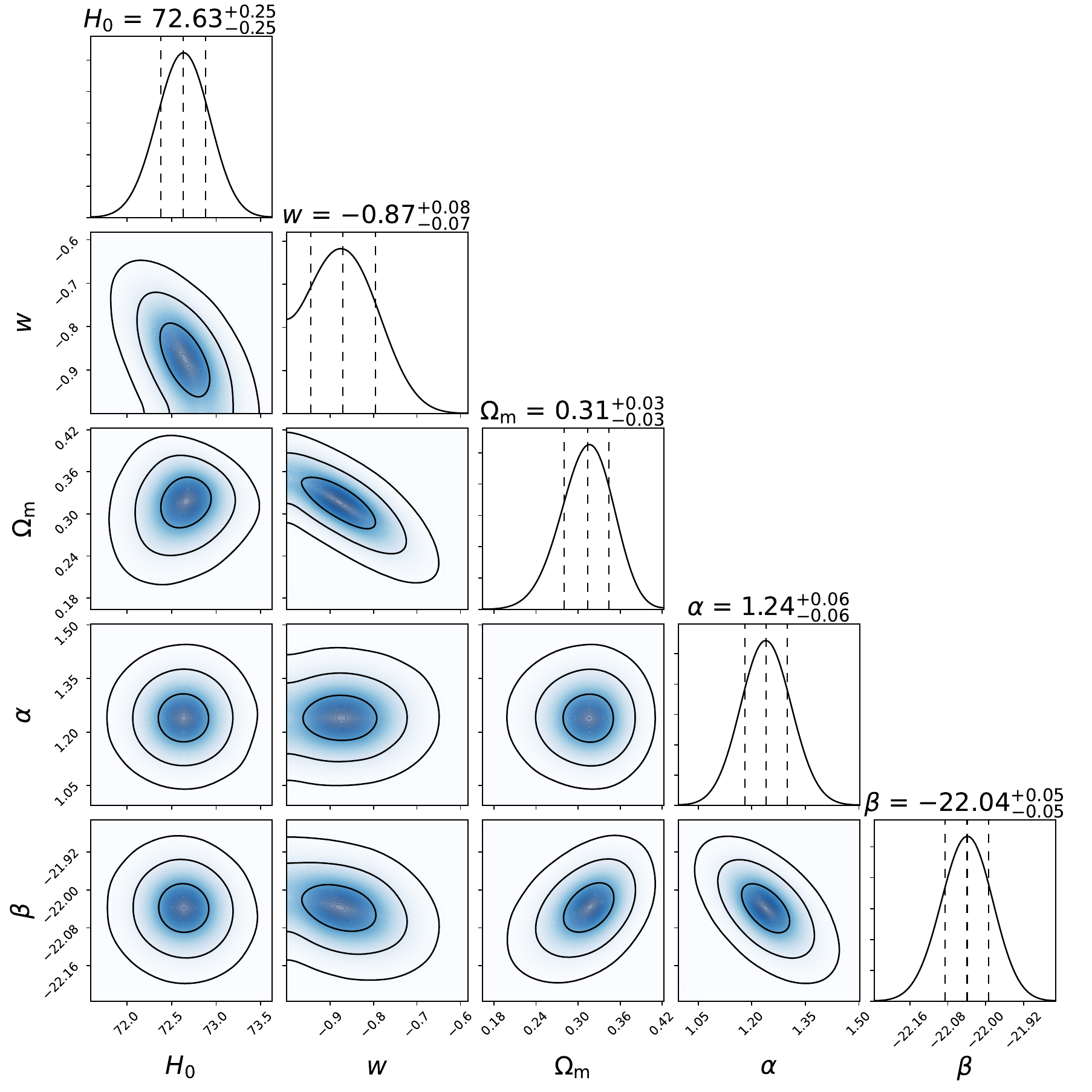}
\caption{Joint confidence level contours of the $w$CDM model parameters inferred from the MCMC-based fitting of the SLSNe and SNe sample. The inner, middle, and outer contours centered at any intersection of parameter pairs indicate the $68\%, 95\%$, and $99.7\%$ confidence levels, respectively.}
\label{F_wCDM_SNe}
\end{figure}

For the flat $w$CDM model, the best-fit results are $\Omega_m = 0.36 \pm 0.07$, $w = -1.81^{+0.83}_{-0.77}$, $\alpha = 1.38 \pm 0.15$ and $\beta = -22.22 \pm 0.14$. After combining the Pantheon+ sample, as presented in Fig. \ref{F_wCDM_SNe}, the best-fit results suggest $H_0 = 72.63 \pm 0.25$ km s$^{-1}$ Mpc$^{-1}$, $\Omega_m = 0.31 \pm 0.03$, $w = -0.87^{+0.08}_{-0.07}$, $\alpha = 1.24 \pm 0.06$ and $\beta = -22.04 \pm 0.05$.

\section{Conclusions}\label{Conclusions}
In this study, combining with the rate estimation from previous works, we revealed a bias in tracing the SFR, manifesting a factor of $(1+z)^{1.2}$. To investigate the capability of SLSNe as standard candles, we further fit the sample with both the peak-decline and peak-colour relations. Our fittings suggest that the peak-colour relation exhibits smaller scattering compared to the peak-decline relation, both appear to be statistically significant. Utilizing the peak-colour relation, we simulated a sample of SLSNe to be observed by CSST, and our estimation suggests that $\approx$360 simulated SLSNe spanning redshifts from $z=0$ to $z=6.5$ will be observed during the 2.5-year 10\,deg$^{2}$ CSST ultra-deep sky survey. The constraints on the flat $\Lambda$CDM model that will be posed by CSST are rather stringent, yielding $\Omega_m = 0.34^{+0.07}_{-0.06}$, $\alpha = 1.40^{+0.15}_{-0.14}$, and $\beta = -22.12 \pm 0.10$. Furthermore, by combining the SLSNe sample with the Pantheon+ sample, we achieved more refined limits on the cosmological parameters.

SLSNe serve as valuable probes bridging the gap between low redshift events and observations of CMB. Their utility as standard candles allows for a deeper exploration of the behavior of our universe and the investigation of dark energy properties at high redshifts. The simulations in our analysis provide a forecast for the cosmological applications of SLSNe in the future CSST $10$ deg$^2$ ultra-deep field survey program. Looking forward to extensive and high-quality observations of SLSNe, we anticipate achieving more precise constraints on cosmological models, with a particular emphasis on advancing our understanding of the dark energy equation of state and Hubble constant \cite{Yu2018}.

\Acknowledgements{
We thank Li shiyu and Wang Xiaofeng for helpful discussion. This work is supported by the National Natural Science Foundation of China (grant Nos. 12273009 and 12393812), the National SKA Program of China (grant No. 2020SKA0120300), the China Manned Spaced Project (CMS-CSST-2021-A12). Jian-Ping Hu is supported by the Jiangsu Funding Program for Excellent Postdoctoral Talent (20220ZB59).
}

\linespread{1}
\begin{table*}
	\centering
	\caption{SLSNe data sample.For each SLSN, their ID name, redshift, peak luminosity, the decline of peak luminosity over $30$ days and the difference of the decline rate between $400$ nm filter and $520$ nm filter over $30$ days are tabulated. The $1 \sigma$ uncertainties in $M(400)_{0}$, $\Delta M(400)_{30}$, and $M(400)_{30}-M(520)_{30}$ are indicated by the numbers in parentheses.}
	\label{SLSNe data}
	\begin{tabular}{ccccc} 
		\hline
		ID & redshift & $M(400)_0$ & $\Delta M(400)_{30}$ & $M(400)_{30}-M(520)_{30}$\\
		\hline
		Gaia16apd & 0.102& -21.87 $\pm$ 0.04 & 0.69 $\pm$ 0.06 & 0.28 $\pm$ 0.07\\
		SN2011ke & 0.143 & -21.23 $\pm$ 0.09 & 0.89 $\pm$ 0.09 & 0.59 $\pm$ 0.03 \\
		SN2012il & 0.175 & -21.54 $\pm$ 0.10 & 1.39 $\pm$ 0.17 & 0.48 $\pm$ 0.13\\
		PTF11rks & 0.190 & -20.61 $\pm$ 0.05 & 0.87 $\pm$ 0.07 & 1.16 $\pm$ 0.15\\
		SN2010gx & 0.230 & -21.73 $\pm$ 0.02 & 0.76 $\pm$ 0.03 & 0.53 $\pm$ 0.03\\
		SN2011kf & 0.245 & -21.74 $\pm$ 0.15 & 0.52 $\pm$ 0.18 & ...\\
		LSQ12dlf & 0.255 & -21.52 $\pm$ 0.03 & 0.76 $\pm$ 0.04 & 0.57 $\pm$ 0.10\\
		LSQ14mo  & 0.256 & -21.04 $\pm$ 0.05 & 1.30 $\pm$ 0.14 & 0.61 $\pm$ 0.02\\
		PTF09cnd & 0.258 & -22.16 $\pm$ 0.08 & 0.71 $\pm$ 0.14 & ...\\
		SN2013dg & 0.265 & -21.35 $\pm$ 0.05 & 1.03 $\pm$ 0.06 & 0.56 $\pm$ 0.10\\
		PS1-10bzj& 0.650 & -21.03 $\pm$ 0.06 & 1.23 $\pm$ 0.32 & 0.94 $\pm$ 0.25\\
		iPTF13ajg& 0.740 & -22.42 $\pm$ 0.07 & 0.19 $\pm$ 0.10 & -0.11 $\pm$ 0.09\\
		DES15X3hm& 0.860 & -21.94 $\pm$ 0.06 & 1.44 $\pm$ 0.07 & 0.11 $\pm$ 0.06\\
		DES17X1amf& 0.920& -21.97 $\pm$ 0.07 & 0.26 $\pm$ 0.15 & 0.08 $\pm$ 0.10\\
		PS1-10ky & 0.956 & -22.05 $\pm$ 0.06 & 0.61 $\pm$ 0.07 & 0.25 $\pm$ 0.06\\
		PS1-11aib & 0.997& -22.05 $\pm$ 0.07 & 0.31 $\pm$ 0.17 & -0.15 $\pm$ 0.07\\
		SCP-06F6 & 1.189 & -22.19 $\pm$ 0.03 & 0.57 $\pm$ 0.15 & ...\\
		PS1-11tt & 1.283 & -21.89 $\pm$ 0.16 & 0.15 $\pm$ 0.20 & ...\\
		PS1-11bam & 1.565 &-22.45 $\pm$ 0.10 & 0.36 $\pm$ 0.14 & ...\\
		DES16C2nm & 1.998 &-22.52 $\pm$ 0.10 & 0.67 $\pm$ 0.11 & ...\\
		
		\hline
	\end{tabular}
\end{table*}

\end{multicols}
\end{document}